\documentclass[conference]{IEEEtran}
\usepackage{times,amsfonts,bbm,dsfont,amsmath,color,amssymb,graphicx,epsfig,multirow,float,algorithm,algorithmic,bm,cite,setspace}
\usepackage[normalem]{ulem}
\usepackage{makecell}
\usepackage{stfloats}
\usepackage{arydshln}
\usepackage{psfrag}
\usepackage{epstopdf}
\usepackage{stfloats}
\usepackage{color,cases}
\usepackage{subfigure}

\begin{document}

\title{A Learning-Based Coexistence Mechanism for LAA-LTE Based HetNets}

\author{Junjie Tan$^{\dag}$, Sa Xiao$^{\dag}$, Shiying Han$^{\ast}$, Ying-Chang Liang$^{\dag}$, {\it Fellow, IEEE} \\
$^{\dag}$University of Electronic Science and Technology of China (UESTC), Chengdu, P. R. China\\
$^{\ast}$Nankai University, Tianjin, P. R. China}


\maketitle
\begin{abstract}
License-assisted access LTE (LAA-LTE) has been proposed to deal with the intense contradiction between tremendous mobile traffic demands and crowded licensed spectrums. In this paper, we investigate the coexistence mechanism for LAA-LTE based heterogenous networks (HetNets). A joint resource allocation and network access problem is considered to maximize the normalized throughput of the unlicensed band while guaranteeing the quality-of-service requirements of incumbent WiFi users. A two-level learning-based framework is proposed to solve the problem by decomposing it into two subproblems. In the master level, a Q-learning based method is developed for the LAA-LTE system to determine the proper transmission time. In the slave one, a game-theory based learning method is adopted by each user to autonomously perform network access. Simulation results demonstrate the effectiveness of the proposed solution.
\end{abstract}

\section{Introduction}
\label{sec:intro}

The exponential growth of mobile devices and the popularity of various mobile applications, like streaming videos, result in at least 7-fold mobile traffic increase by 2021\cite{cisco}. To accommodate such terrific mobile traffic via wireless access, the rarity of spectrum resource has become a main bottleneck for further improvement in the system capacity \cite{RN17}. Therefore, how to broaden the available spectrum has been considered as a major challenge in the future wireless systems by both academia and industry.

Introducing LTE systems to use the unlicensed bands currently occupied by WiFi system is one of the efficient ways to cope with the challenge of spectrum scarcity. The corresponding standard called \emph{licensed-assisted access} (LAA) has been developed by \emph{3rd Generation Partnership Project} (3GPP) since 2014 \cite{nielsen2014lte}. In LAA-LTE systems, LTE users are allowed to occupy the unlicensed bands for data transmission. However, since the \emph{distributed coordination function} (DCF) and contention-based MAC protocols, e.g. CSMA, are employed, the performance of the WiFi system can be severely degraded if aggressive spectrum sharing strategies are adopted by LTE users \cite{RN78}. Therefore, efficient and fair coexistence mechanisms to maximize the usage of unlicensed bands while maintaining the \emph{quality-of-service} (QoS) requirements of WiFi users should be designed for LAA-LTE systems.

Thanks to its WiFi-friendly nature and the regulatory requirement of certain countries, \emph{listen-before-talk} (LBT) is widely used for the coexistence between the WiFi and LAA-LTE systems\cite{RN78}. There has been some preliminary work regarding the LBT-based LAA-LTE systems. In \cite{RN173}, a contention-window optimization method has been proposed to maximize the throughput of the LAA-LTE system, while in \cite{RN17}, joint routing selection and resource allocation algorithms have been developed for both real-time and non-real-time applications in LAA-LTE \emph{heterogeneous networks} (HetNets). In \cite{RN6}, a novel LBT-based MAC protocol has been designed to maximize the normalized throughput of unlicensed bands without sacrificing the performance of incumbent WiFi users. Note that all aforementioned work only considers the performance analysis and parameter optimization, and does not take network access into consideration. In \cite{RN172}, joint resource allocation and network access has been investigated to minimize the collision probability of the WiFi system. However, this work requires a central controller to schedule the activities of each user, therefore may neither be scalable especially when the number of users is large, nor be adaptable to the variation of the network settings.

Motivated by the above work, in this paper, we develop a learning-based two-level mechanism for the coexistence in LAA-LTE based HetNets, which operates in a distributed manner and jointly solves the resource allocation and network access problem with the objective to maximize of the normalized throughput of the unlicensed bands.
In the master level, a Q-learning based method is developed for the LAA-LTE \emph{base station} (BS) to determine the optimal transmission time in the unlicensed bands. In the slave one, a game-theory-based learning method is adopted by each user to autonomously choose the proper network to access. Simulation results show that the proposed method is not only effective and efficient, but also adaptable to the variational network settings.

The rest of the paper is organized as follows. In Section \ref{sec:sysmodel}, the system model is described, which is followed by the throughput analysis and problem formulation in Section III. To efficiently solve the problem, we propose a two-level learning-based framework in Section IV. Then we present simulation results in Section V. Finally, conclusions are drawn in Section VI.

\section{System Model}
\label{sec:sysmodel}
\subsection{LAA-LTE based HetNets}

\begin{figure}
[htbp]
\centering
  \includegraphics[width=.8\columnwidth]{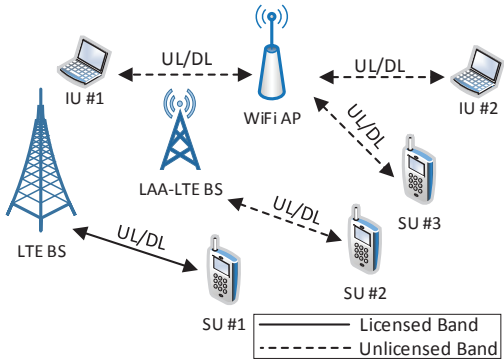}
  \caption{The system model for a LAA-LTE based HetNet.}\label{sysmodel1}
  \vspace{-10pt}
\end{figure}

In this paper, we consider a LAA-LTE based HetNet as shown in Fig.\ref{sysmodel1}, where the LTE network operates in the licensed bands, and the LAA-LTE and WiFi networks share the same unlicensed band by transmitting in different fractions of time. In the system, there are $N_1$ \emph{incumbent users} (IUs) and $N_2$ \emph{smart users} (SUs). Specially, the IUs are all associated to the WiFi network while the SUs are equipped with multi-\emph{radio access technologies} (RATs) to access any of the three networks. The data traffic of the IUs and SUs is assumed to follow Poisson process and different users may have various packet arrival rates due to distinct traffic demands. For analytical simplicity, we consider a basic scenario including one WiFi \emph{access point} (AP) and one LAA-LTE BS in the WiFi and LAA-LTE networks, respectively. In addition, as we only focus on the performance of the unlicensed band, we also assume the LTE network has sufficient resource and can provide reliable supports to SUs especially when the unlicensed band is crowded. In the remaining parts of the paper, we define $\mathcal{I}$ and $\bm{\lambda}_{1}=\{\lambda_{1,1},...,\lambda_{1,N_1}\}$ as the user set and the packet arrival rate set of IUs, respectively, where $\lambda_{1,i}$ is the average packet arrival rate of IU $i$ per packet transmission time $T$. Similarly, the user set and packet arrival rate set of the SUs can be defined as $\mathcal{S}$ and $\bm{\lambda}_{2}=\{\lambda_{2,1},...,\lambda_{2,N_2}\}$, respectively.

\subsection{Protocol Description}

\begin{figure}
[t]
\centering
  \includegraphics[width=0.9\columnwidth]{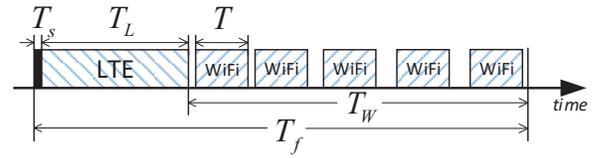}
  \caption{The MAC protocol of LAA-LTE.}\label{mac}
  \vspace{-10pt}
\end{figure}

To prevent the LAA-LTE network from interrupting the ongoing transmission in the WiFi network, the frame-based LBT protocol mentioned in \cite{RN6} is adopted in this paper. The LAA-LTE network with LBT mechanism transmits for a certain period of time once the channel is sensed to be idle. On the other hand, the WiFi network adopts 1-persistent CSMA protocol and therefore can only transmit when the LAA-LTE transmission phase ends. The frame structure of the protocol is illustrated in Fig. \ref{mac}
, where the total frame duration, sensing time, LAA-LTE transmission time, and WiFi transmission time are denoted as $T_f$, $T_s$, $T_L$, and $T_W$, respectively. As $T_s$ is relatively small than $T_L$, we can ignore $T_s$ and then have $T_f=T_L+T_W$. In addition, for expressional simplicity, the frame duration, LAA-LTE transmission time, and WiFi transmission time can be normalized over per packet transmission time $T$, which results in  $\theta=T_f/T$, $\beta=T_L/T$, and $\gamma=T_W/T$.

\begin{figure*}[htp]
\begin{align}
B(\mathcal{S}_1,\beta) &= {z^{ - 1}} + \frac{{{\beta ^2}}}{{2\theta }}\left[ {\frac{{1 + {{(1 - z)}^\gamma }}}{{1 - {{(1 - z)}^\gamma }}}} \right] + \frac{1}{{2\theta z}}\left\{ {\sigma \gamma z + 2\beta  + (1 - \sigma ){{\left[ {1 - (1 - z)} \right]}^\gamma }} \right\}, \label{B}\\
U(\mathcal{S}_1,\beta) &= {\theta ^{ - 1}}\left[ {{G_1}{{\left( {1 - z} \right)}^{ - 1 + \gamma }}{z^{1 + \beta }}\left( {\frac{1}{z} + \frac{\beta }{{1 - {{\left( {1 - z} \right)}^\gamma }}}} \right)\left( {1 + \gamma } \right)  } \right. +{G_1}\left( {\theta  - 1 - \frac{{{{\left( {1 - z} \right)}^{ - 1 + \gamma }}z\beta }}{{1 - {{\left( {1 - z} \right)}^\gamma }}}} \right) \nonumber \\
& + \frac{{1 - {z^{1 + \beta }}}}{{{G_1}}} + \left. {\left( {1 - {{\left( {1 - z} \right)}^\gamma }} \right){z^\beta }\left( { - z + \left( {1 - z} \right)\left( {1 + \frac{1}{{{G_1}}} + \beta } \right)} \right) - \frac{{\left( { - 1 + {z^\sigma }} \right)\left( { - 1 + \gamma } \right)}}{{{G_1}\sigma }}} \right]. \label{U}
\end{align}
\hrulefill
\vspace{-5pt}
\end{figure*}
\section{Throughput Analysis and Problem Formulation}\label{sec:ProblemFormulation}
In this section, we first analyze the normalized throughput, i.e. the successful transmission time ratio, of the unlicensed band, and then formulate a joint resource allocation and network access problem to maximize the overall normalized throughput. In the remaining of the paper, the term of throughput stands for the normalized throughput.

\subsection{Throughput Analysis}

Let $\mathcal{S}_1$, $\mathcal{S}_2$ and $\mathcal{S}_3$ be the sets of SUs staying in the WiFi, LAA-LTE, and LTE networks, respectively, where $\mathcal{S}_1\cup\mathcal{S}_2\cup\mathcal{S}_3=\mathcal{S}$ and $\mathcal{S}_1\cap\mathcal{S}_2\cap\mathcal{S}_3=\emptyset$. The throughput of the unlicensed band, consisting of the throughput of WiFi network and LAA-LTE network, can be expressed as follows:

\subsubsection{WiFi Network}

Because of the DCF, there exists packet collisions in the WiFi network. Therefore, the throughput of the WiFi network should be the ratio of the successful transmission time to the whole frame duration. By extending the results in \cite{RN6} and \cite{RN70}, the throughput can be expressed as

\begin{equation}\label{eq:wifirate}
{R_W}({{\cal S}_1},\beta ) = \frac{{U({{\cal S}_1},\beta )}}{{B({{\cal S}_1},\beta ) + 1/{G_1}}},
\end{equation}
where ${G_1}{\rm{ = }}\sum\nolimits_{i \in I} {{\lambda _{1,i}}} {\rm{ + }}\sum\nolimits_{j \in {S_1}} {{\lambda _{2,j}}}$ is the total average data traffic of the WiFi network, and $B(\mathcal{S}_1,\beta)$ and $U(\mathcal{S}_1,\beta)$ are respectively given by (\ref{B}) and (\ref{U}) at the top of next page, with $\sigma$ denoting the length of a mini-slot that the time is discretized with. $B(\mathcal{S}_1,\beta)$, $U(\mathcal{S}_1,\beta)$ and $1/{G_1}$ are actually the expected busy, non-collision and idle channel duration, respectively.

\subsubsection{LAA-LTE Network}

Thanks to the centralized coordination for data transmission, there are no packet collisions in the LAA-LTE network. Therefore, the throughput is exactly the ratio of the transmission time to the whole frame duration. Let $G_2=\sum\nolimits_{j \in {S_2}} {{\lambda _{2,j}}}$ be the total average data traffic of $\mathcal{S}_2$. When the LAA-LTE network is saturated, i.e. $\beta \le \theta G_2$, the transmission time of the LAA-LTE network is $\beta$ and the corresponding throughput is $\beta/\theta$. On the other hand, when the LAA-LTE network is unsaturated, i.e. $\beta >\theta G_2$, the transmission time is $\theta G_2$ and the throughput is $G_2$. Therefore, we have
\begin{equation}\label{eq:laarate}
R_{LAA}(\mathcal{S}_2,\beta)=\min(\beta/\theta,G_2).
\end{equation}

Based on \eqref{eq:wifirate} and \eqref{eq:laarate}, the total throughput of the unlicensed band can be written by
\begin{equation}\label{eq:sumrate}
R_{t}(\mathcal{S}_1,\mathcal{S}_2,\beta)=R_{W}(\mathcal{S}_1,\beta)+R_{LAA}(\mathcal{S}_2,\beta).
\end{equation}
\subsection{Problem Formulation}
To achieve the fair coexistence, we enforce following two constraints for the throughput of IUs and SUs.

First, if SUs are allowed to access the LAA-LTE or LTE network, the throughput of IUs under this scenario should not be worse than that can be achieved when all SUs access the WiFi network. That is

\begin{equation}\label{eq:cons1}
\frac{{{G_3}{R_W}({{\cal S}_1},\beta )}}{{{G_1}}} \ge \frac{{{G_3}{R_0}({\cal S})}}{{{G_4}}},
\end{equation}
where $G_3=\sum\nolimits_{i \in I} {{\lambda _{1,i}}}$ is the total average data traffic of IUs, ${G_4}{\rm{ = }}\sum\nolimits_{i \in I} {{\lambda _{1,i}}} {\rm{ + }}\sum\nolimits_{j \in S} {{\lambda _{2,j}}} $ is the total average data traffic of the pure WiFi network with $\mathcal{S}_1=\mathcal{S}$ and $\beta=0$, and $R_0$ is the throughput of the pure WiFi network, which is given by
\begin{equation}\label{eq:purerate}
{R_0} = \frac{{{G_4}(1 + {G_4}){e^{ - {G_4}}}}}{{{G_4} + {e^{ - {G_4}}}}}.
\end{equation}

Second, intuitively, if SUs want to access the LAA-LTE network, the obtained throughput of them should be higher than that can be achieved in the pure WiFi network, i.e.
\begin{equation}\label{eq:cons2}
{R_{LAA}}({{\cal S}_2},\beta ) \ge \frac{{{G_2}{R_0}}}{{{G_4}}}.
\end{equation}

With the constraints given by \eqref{eq:cons1} and \eqref{eq:cons2}, the throughput maximization problem can be formulated as follows.

\textbf{\underline{Problem 1:}}
\begin{eqnarray}
\max_{\{\beta,\mathcal{S}_1,\mathcal{S}_2,\mathcal{S}_3\}}&& R_{t}(\mathcal{S}_1,\mathcal{S}_2,\beta)
\label{eq:P1}\nonumber\\
\text{s.t}.~~~~~&& (\ref{eq:cons1}), (\ref{eq:cons2}), \nonumber
\label{eq:P1C1} \\
&& \mathcal{S}_1\cup\mathcal{S}_2\cup\mathcal{S}_3=\mathcal{S},\\
&& \mathcal{S}_1\cap\mathcal{S}_2\cap\mathcal{S}_3=\emptyset.
\end{eqnarray}

Since $R_{t}(\mathcal{S}_1,\mathcal{S}_2,\beta)$ is a unimodal function of $\beta$ for any given $\mathcal{S}_1$ and $\mathcal{S}_2$ \cite{RN6}, the above problem can be optimally solved in two steps. First, the optimal $\beta^*$ is determined for all the possible combinations of $\left\{ {{{\cal S}_1},{{\cal S}_2},{{\cal S}_3}} \right\}$ by using the method mentioned in \cite{RN6}. Then, the $\left\{ {{{\cal S}_1^*},{{\cal S}_2^*},{{\cal S}_3^*}} \right\}$ rendering the highest $R_t$ is chosen as the optimal network access strategy.
However, this optimal solution has the computational complexity of $O(3^{N_2})$, which is prohibitively high especially when $N_2$ is large. What's more, the algorithm is centralized and needs to be rerun once the network setting changes, e.g. a new SU arrives. To deal with these issues, a distributed learning-based mechanism is proposed in the next section, which not only
has approximate performance of the optimal solution, but also comes with much lower computational complexity and more adaptability.

\section{A Learning-based Mechanism}\label{sec:DisProb}

In this section, we develop a learning-based mechanism to solve {\bf{Problem~1}}. We first introduce the framework of the proposed mechanism, which decouples the problem into \emph{distributed network access} (DNA) and \emph{resource allocation} (RA) subproblems, and then propose learning-based algorithms to solve the subproblems.

\subsection{Two-Level Intelligent Resource Allocation and Distributed Network Access Framework}
\begin{figure}
[!b]
\vspace{-5pt}
\centering
\vspace{-5pt}
\subfigure[The information and action flowchart.]
{
\centering
\label{flowchart}
\includegraphics[width=1\columnwidth]{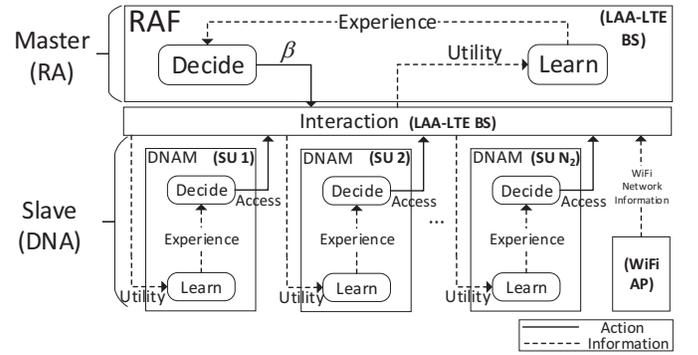}
}
\subfigure[The procedure of the proposed algorithms.]
{
\centering
\label{procedure}
\includegraphics[width=1\columnwidth]{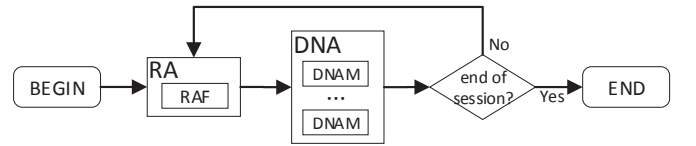}
}
\caption{The two-level learning-based framework.}
\label{framework}
\end{figure}
The framework of the proposed two-level learning-based mechanism is illustrated in Fig. \ref{framework}. In the master level, the \emph{resource allocation function} (RAF) is employed in the LAA-LTE BS to allocate appropriate time resource for the LAA-LTE network. In the slave level, the \emph{distributed network access module} (DNAM) is implemented in each SU to autonomously choose network to access. Both the RAF and DNAM operate based on close-loop learning methods, thus can enhance themselves from the knowledge of experienced utilities.


As shown in Fig. \ref{flowchart}, there exists cooperation among different entities. The LAA-LTE BS needs to first gather the choice of each SU and the information of WiFi network to compute the instantaneous utility, and then broadcast the utility to SUs. The utility serves as the learning experience for the RAF and DNAM to refine their actions. Since the limited amount of broadcast information is required, the proposed algorithm causes low signaling overhead.
Fig. \ref{procedure} describes the whole procedure of the proposed algorithms, where RA and DNA are successively executed until the end of the session, i.e. no SUs exist.

\subsection{Distributed Network Access}\label{sec:AUA}

For a fixed $\beta$, the DNAM intends to allow each SU to perform network access distributively and autonomously. Therefore, the behavior of SUs can be analyzed from the perspective of game theory.

Let $a_j$ denote the action of SU $j$, where $a_j=1$, $a_j=2$, and $a_j=3$ represent the choices for the WiFi, LAA-LTE, and LTE networks, respectively. Then the utility function of SU $j$ can be stated as
\begin{align}\label{eq:utility}
&{u_j}({a_j},{\bm{a}_{ - j}}) \notag\\
&=\left\{ {\begin{array}{*{20}{l}}
{0,}&{{\textmd{if}}~\eqref{eq:cons1}~\textmd{or}~\eqref{eq:cons2}~\textmd{is~not~satisfied.}}\\
{{R_t}({{\cal S}_1},{{\cal S}_2},\beta ),}&{{\textmd{o.w.}}}
\end{array}} \right.
\end{align}
where ${{\bm{a}}_{ - j}} = \left\{ {{a_1},...,{a_{j - 1}},{a_{j + 1}},...,{a_{{N_2}}}} \right\}$ is the joint choices of SUs excluding SU $j$. Notice that the utility functions of different SUs are identical, i.e., $u(\bm{a})={u_j}({a_j},{{\bm{a}}_{ - j}}),\forall j\in \mathcal{S}$, where ${\bm{a}} = \left\{ {{a_1},...,{a_{{N_2}}}} \right\}$. Therefore, the behavior of SUs driven by $u(\bm{a})$ can be modeled as a \emph{common interest game} $\cal{G}$$=[\mathcal{S},\mathcal{A},u(\bm{a})]$.
According to \cite{la2016potential}, as a special case of potential games, the common interest game $\cal{G}$ exists at least one pure \emph{Nash equilibrium} (NE) $\bm{a}^*=\left\{{a^*_1},...,a^*_{N_2}\right\}$ satisfying
\begin{equation}
u(a_j^*,\bm{a}_{ - j}^*) \ge u({a'_j},\bm{a}_{ - j}^*),~\forall {a'_j} \ne a_j^*,a' \in {\mathcal{A}_j,j\in\mathcal{S}},
\end{equation}
and $\bm{a}^*$ is also a maximizer for the utility function in \eqref{eq:utility}.

There are some methods, like fictitious play and best response dynamics, to effectively achieve NEs. However, they need each user to have the knowledge of the actions of other users, which may cause heavy signaling overhead. To avoid the signaling among SUs, a \emph{stochastic learning} (SL) method is adopted in the DNAM and its whole procedure is listed in \textbf{Algorithm 1}.
\begin{algorithm}[!thp]
\caption{The SL Method in The DNAM}
{\normalsize
\begin{algorithmic}[1]
\STATE \textbf{Initialize:} $\bm{p}_j(0)=\{1/3,1/3,1/3\}$, $n=0$.
\REPEAT
\STATE Choose an action ${a}_j(n)$ according to $\bm{p}_j(n)$.
\STATE Act with ${a}_j(n)$ and obtain ${u}(n)$.
\STATE Update $\bm{p}_j(n+1)$ by \eqref{DNAMupdate},
\begin{align} \label{DNAMupdate}
&{p_{j,k}}(n + 1)\notag\\
&=\left\{ {\begin{array}{*{20}{l}}
{{p_{j,k}}(n) - {\kappa _j}{u_j}(n){p_{j,k}}(n),}&{{\textmd{if }}~k = {{a}_j}(n)}\\
{{p_{j,k}}(n) + {\kappa _j}{u_j}(n)(1 - {p_{j,k}}(n)),}&{{\textmd{o}}{\textmd{.w}}{\rm{.}}}
\end{array}} \right.
\end{align}
\STATE $n=n+1$.
\UNTIL {\eqref{eq:cond1} or \eqref{eq:cond2} is satisfied}
\begin{numcases}{}
|\mathop {\max }\limits_k {p_{j,k}}(n) - 1| \le \varepsilon \label{eq:cond1}\\
n\ge n_{\max}~{\textmd{and}}~{\left\| {{\bm{p}_j}(n) - \bm{p}_j(0)} \right\|_2} \le \varepsilon \label{eq:cond2}
\end{numcases}
\end{algorithmic}}
\end{algorithm}

As shown in \textbf{Algorithm 1}, the algorithm starts with an equal mixed strategy $\bm{p}_j(0)=\{p_{j,1}(0),p_{j,2}(0),p_{j,3}(0)\}=\{1/3,1/3,1/3\}$. $p_{j,k}(n)$ denotes the probability of SU $j$ taking action $k$ at $n$-th iteration. At $n$-th iteration, an action $a_j(n)$ is determined according to $\bm{p}_j(n)$ and the instantaneous utility $u(n) = u(\bm{a}(n)) = u_j(\bm{a}(n))$ is obtained from the broadcasted information of the LAA-LTE BS by \eqref{eq:utility}, where $\bm{a}(n) = \{ a_1(n), a_2(n), ..., a_{N_2}(n)\}$. After that, the mixed action profile at next iteration $\bm{p}_j(n+1)$ is updated with the given value of $u(n)$ and step size $\kappa _j$ with \eqref{DNAMupdate}, according to \cite{RN128}. Finally, the loop ends until one of the stop conditions in \eqref{eq:cond1} and \eqref{eq:cond2} is met.

Note that ${R_W}({{\cal S}_1},\beta )$ is a decreasing function of $\beta$ and it satisfies ${R_W}({S_1},\theta) = 0$. Therefore, there must exist ${\beta}_{\max}$ such that \eqref{eq:cons1} is violated for any combination of $\left\{ {{{\cal S}_1},{{\cal S}_2},{{\cal S}_3}} \right\}$ when $\beta>{\beta}_{\max}$. For those $\beta$ satisfying $\beta>{\beta}_{\max}$, the utility function in \eqref{eq:utility} always returns zero, which makes $\bm{p}_j(n)=\bm{p}_j(0)$ until the maximum number of iterations $n_{\max}$ is reached. In this case, \eqref{eq:cond2} is activated and a new $\beta$ is required from the master level for the future operation. For those feasible $\beta$ satisfying $\beta\le{\beta}_{\max}$, \eqref{eq:cons1} can be satisfied by some combinations of $\left\{ {{{\cal S}_1},{{\cal S}_2},{{\cal S}_3}} \right\}$. In this case, the algorithm is guaranteed to converge to a pure NE according to \cite{RN128} and \eqref{eq:cond1} is thus met.

Because of the nonconvexity and noncontinuity of utility function (\ref{eq:utility}), most of the pure NEs are not the global optimal points. However, the simulation results still show that the SL algorithm has approximate performance of global maximizers.

\subsection{Resource Allocation}\label{sec:RL}
Because of the non-uniqueness and local optimality of NEs, the DNAMs may obtain different utilities for a given $\beta$. To achieve better long-term performance as well as being adaptable to the variational network settings, we introduce a Q-learning based method to make decisions based on experience and historical rewards.

Standard Q-learning is usually used for \emph{Markov decision process} (MDP) \cite{sutton1998reinforcement}, which requires a direct relationship between the actions, i.e. the discretized LAA-LTE transmission time $\mathcal{A}_L=\{\beta_1,\beta_2,...,\theta\}$ and states, i.e. the network settings $\{\mathcal{I},\mathcal{S}_1,\mathcal{S}_2,\mathcal{S}_3\}$. However, since a clear connection between the change of network settings and the choice of $\beta$ is hard to be found, the RA problem can hardly be modeled as a MDP. Therefore, we turn to a state-free Q-learning method, known as \emph{stateless Q-learning} (SLQL) \cite{kapetanakis2002reinforcement}, to solve the RA problem.

The traditional SLQL algorithm mainly composes of two steps, namely the Q-value update step and the action selection step. In the first step, the Q-value of a chosen $\beta$, which is the estimated utility of $\beta$ and denoted by $Q(\beta)$, is updated according to the following rule,
\begin{equation}\label{eq:update}
Q(\beta)=Q(\beta)+\alpha(r-Q(\beta)),
\end{equation}
where $r$ is the received reward, which equals the value of the utility function in (\ref{eq:utility}) after DNA is completed.
Note that the update rule in (\ref{eq:update}) implies that the information of historical rewards are partly stored with the help of the update factor $\alpha$, which can also help to smooth the impacts of different NEs.

In the action selection step, the RAF takes either the exploration or exploitation mode to select $\beta$. The exploration mode aims to collect enough experience for a better decision, and thus the $\beta$ is randomly selected from the action set $\mathcal{A}_L$. On the other hand, in the exploitation mode, the RAF insists on the best action known so far, therefore the $\beta$ rendering the highest $Q(\beta)$ is selected. The tradeoff between the exploration and exploitation modes is determined by a probability factor $\omega$. Specially, if $\omega$ is large, the exploration mode is more preferred than the exploitation one, and otherwise, the converse is true.

Unfortunately, due to the random selection in the exploration mode, the traditional SLQL algorithm may frequently choose those infeasible $\beta$, i.e., $\beta>{\beta}_{\max}$, which induces severe performance loss. Therefore, an \emph{enhanced SLQL} algorithm is proposed to reduce the selections of the infeasible $\beta$ by restricting the action sets in both the exploration and exploitation modes. The steps of the eSLQL algorithm for the RA problem are summarized in \textbf{Algorithm 2}. Specially, if this is the first run of the algorithm, i.e., $\beta'$ does not exist, the RAF goes through the following steps for initialization:
\begin{itemize}
  \item Set the initial values of $Q(\beta)$ and $\omega$;
  \item Find the value of the threshold $\beta_{\max}$ by using bisection search over $\mathcal{A}_L$, and then determine the feasible action set $\mathcal{A}_F$, infeasible action set $\mathcal{A}_I$, and trial set $\mathcal{A}_T$ based on $\beta_{\max}$, where $\mathcal{A}_T$ is a subset of $\mathcal{A}_I$ and its size is called exploration factor $\delta$, i.e. $|\mathcal{A}_T|=\delta$;
  \item Choose an initial action $\beta'$ from $\mathcal{A}_F$.
\end{itemize}

If not, the RAF takes the following steps to find $\beta$ based on the reward $r$ received after the end of DNA:
\begin{itemize}
  \item Update $Q(\beta)$ according to $r$ and \eqref{eq:update};
  \item Update $\mathcal{A}_F$, $\mathcal{A}_T$ and $Q(\beta)$ if the feasibility of the action $\beta'$ changes. Specially, lines 7 and 8 correspond to the case that a feasible action $\beta'$ becomes infeasible, while lines 9 and 10 are operated when an infeasible action $\beta'$ becomes feasible;
  \item Take either the exploration mode (line 13) or the exploitation mode (line 15) to update $\beta'$ according to $\omega$, where $rand()$ generates a random number in $[0,1]$.
\end{itemize}

\begin{algorithm}[!thp]
\caption{The eSLQL Algorithm in The RAF}
{\normalsize
\begin{algorithmic}[1]
\IF {$\beta'$ does not exist}
\STATE Set $Q(\beta_i)=0,\forall i \in \mathcal{A}_L$, $\omega\in(0,1)$.
\STATE Determine $\beta_{\max}$, set $\mathcal{A}_F=\{\beta|\beta\in\mathcal{A}_L, \beta\le\beta_{\max}\}$, $\mathcal{A}_I=\{\beta|\beta\in\mathcal{A}_L,\beta>\beta_{\max}\}$, $\mathcal{A}_T\subseteq \mathcal{A}_I$.
\STATE Randomly choose $\beta'\in\mathcal{A}_F$.
\ELSE
\STATE Obtain reward $r$ and update $Q(\beta')$ with (\ref{eq:update}).
\IF {$\beta'\in\mathcal{A}_F$ and $r=0$}
\STATE Set $\mathcal{A}_F=\{\beta|\beta\in\mathcal{A}_L, \beta<\beta'\}$, $\mathcal{A}_I=\{\beta|\beta\in\mathcal{A}_L, \beta\ge\beta'\}$, $\mathcal{A}_T\subseteq \mathcal{A}_I$ and $Q(\beta_i)=0,\forall i \in \mathcal{A}_L$.
\ELSIF {$\beta_{t}\in\mathcal{A}_T$ and $r\ne0$}
\STATE Set $\mathcal{A}_F=\{\beta|\beta\in\mathcal{A}_L, \beta\le\beta'\}$, $\mathcal{A}_I=\{\beta|\beta\in\mathcal{A}_L, \beta>\beta'\}$, $\mathcal{A}_T\subseteq \mathcal{A}_I$ and $Q(\beta_i)=0,\forall i \in \mathcal{A}_L$.
\ENDIF
\IF {$rand()$$<\omega$}
\STATE Randomly choose $\beta'$ from $\mathcal{A}_F \cup \mathcal{A}_T$.
\ELSE
\STATE Choose $\beta'$ from $\mathcal{A}_F$ with the biggest Q-value.
\ENDIF
\ENDIF
\end{algorithmic}}
\vspace{-2pt}
\end{algorithm}

\section{Simulation Results}

In this section, we evaluate the performance of the proposed algorithms. Packet transmission time $T$, frame duration of LAA-LTE network $T_f$ and mini-slot length $\sigma$ are chosen as $10ms$, $300ms$ and $20\mu s$ respectively.
\subsection{The SL algorithm}
In this part, the performance of the SL algorithm is evaluated under the scenario where $\beta= 1.618$, $N_1=5$, $N_2=6$, $\bm{\lambda}_1=\{$0.03, 0.05, 0.08, 0.09, 0.11$\}$, and $\bm{\lambda}_2=\{$0.05, 0.03, 0.05, 0.3, 0.02, 0.1$\}$. For comparison, we use the exhaustive search method to deal with the network access problem and the corresponding optimal throughput is $0.4754$. Table I illustrates the performance of the top 15 most frequently reached NEs in a 100000 \emph{Monte-Carlo} (MC) test of the SL algorithm, where the throughput and the appearance frequency are listed in the last two columns. In the table, it is evident that all the NEs achieve more than $95\%$ performance of the optimum and two of them (marked with $*$) are exactly the optimal solutions. As these NEs are achieved with relatively high probability and the average throughput of the MC test is 0.4552, the effectiveness of the algorithm can be demonstrated. In addition, from the table, it can be observed that the heavily high traffic user, i.e. SU $4$, prefers LTE network because its existence on unlicensed band will induce heavy utility decrease even if the choice of LTE gets zero payoff intuitively. Also, for slightly high traffic user, i.e. SU $6$, prefers LAA-LTE network rather than WiFi network to boost the overall performance of the unlicensed band by avoiding contention. It is worth noticing that though shown with a specific scenario because of the limited space, the phenomena are generalizable with other setups.

\begin{figure}[t]
\vspace{-5pt}
    \begin{minipage}[b]{0.49\linewidth}
        \centering
        \includegraphics[width=4.0cm]{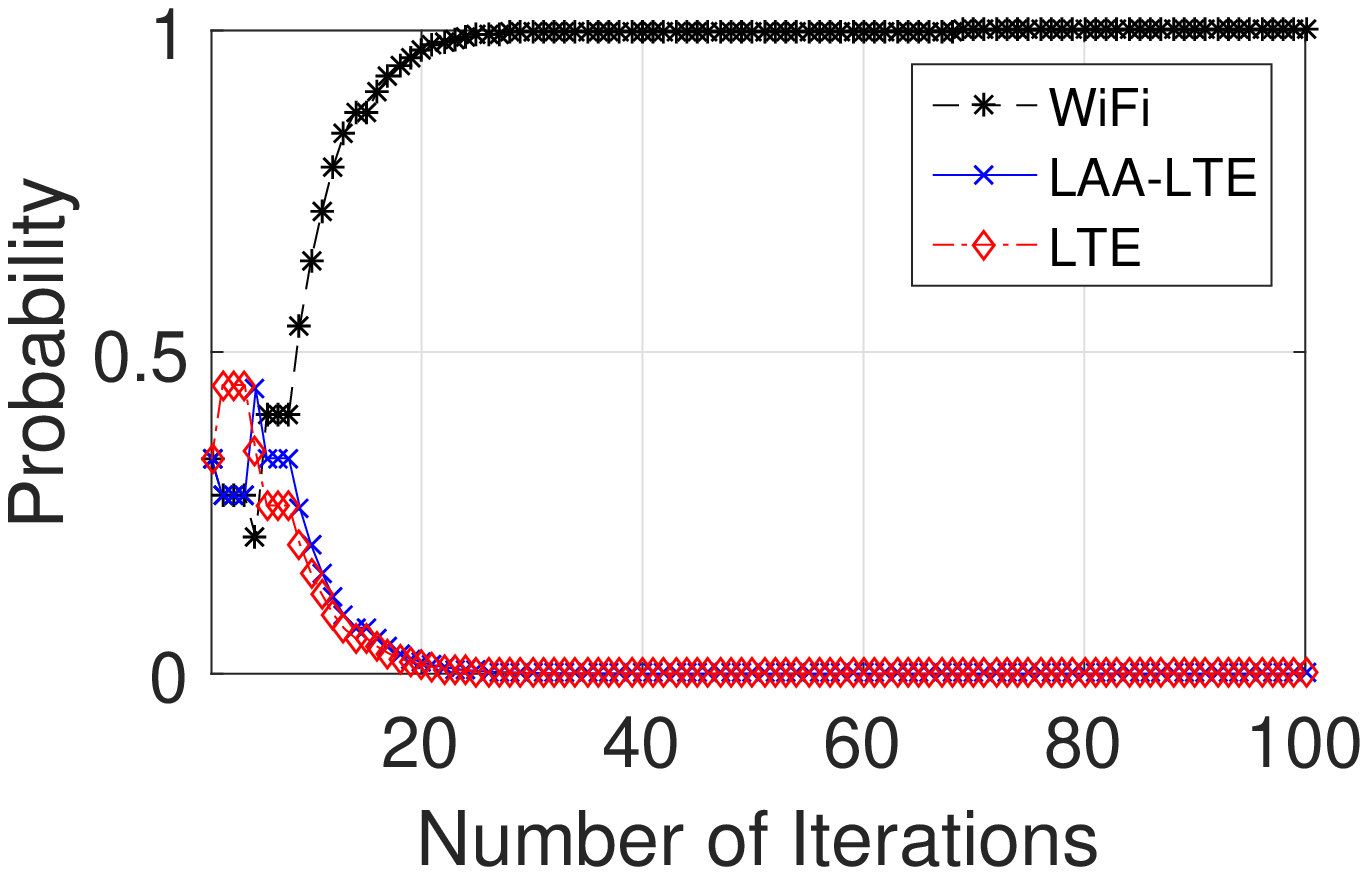}
    \end{minipage}%
    \begin{minipage}[b]{0.49\linewidth}
        \centering
        \includegraphics[width=4.0cm]{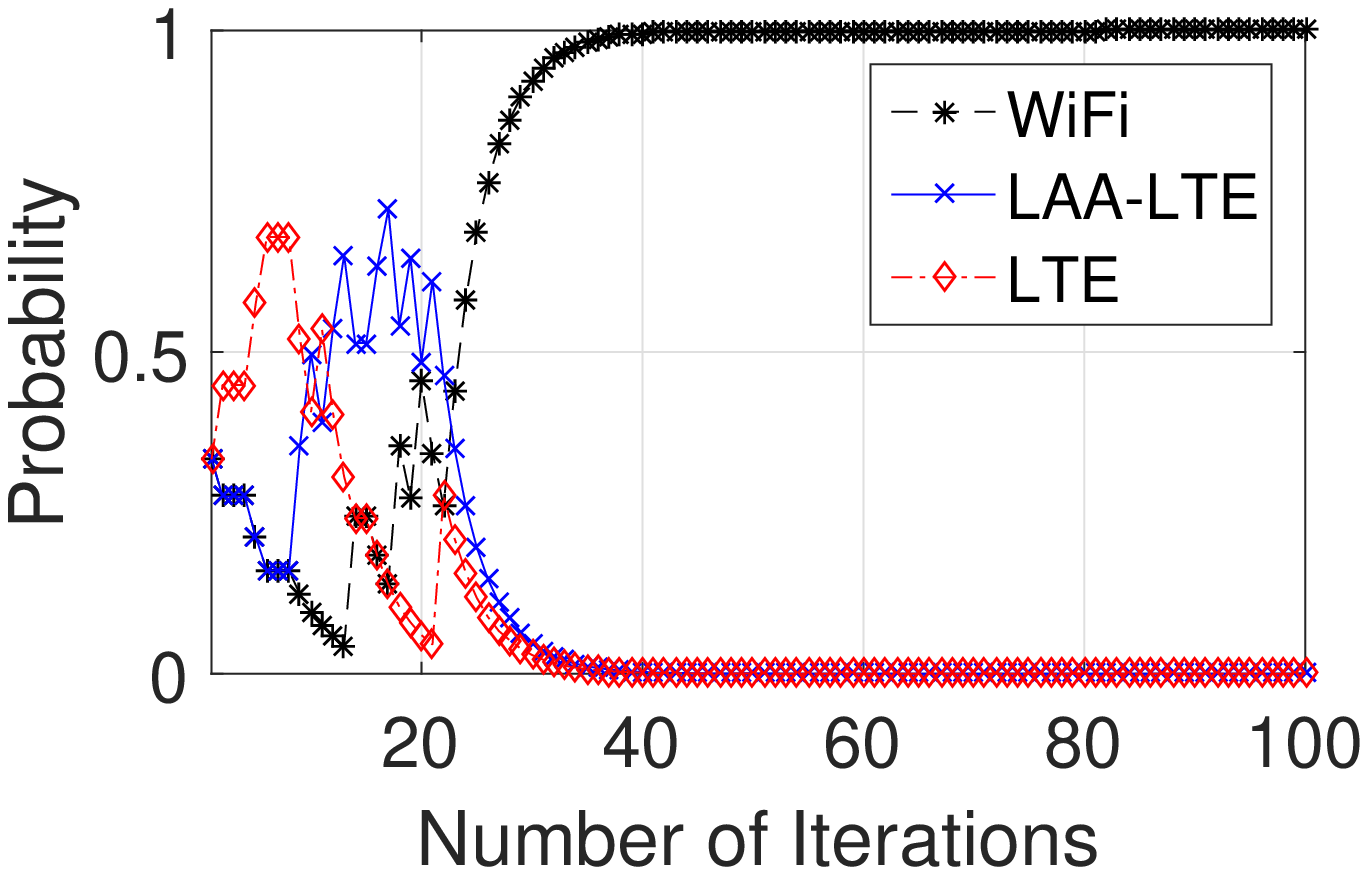}
    \end{minipage} \\[0.90mm]
    \begin{minipage}[b]{0.49\linewidth}
        \centering
        \includegraphics[width=4.0cm]{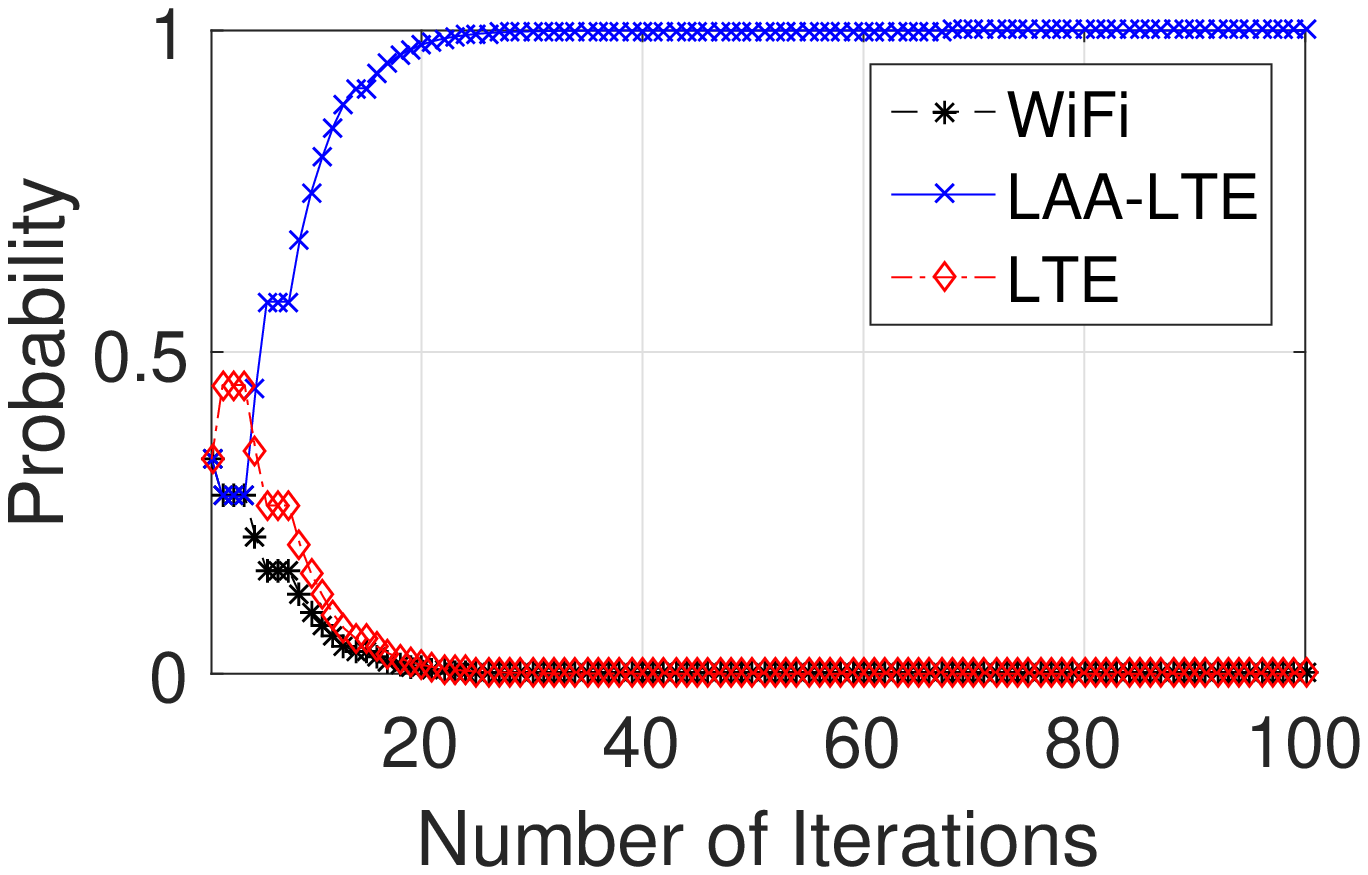}
    \end{minipage}
    \begin{minipage}[b]{0.45\linewidth}
        \centering
        \includegraphics[width=4.0cm]{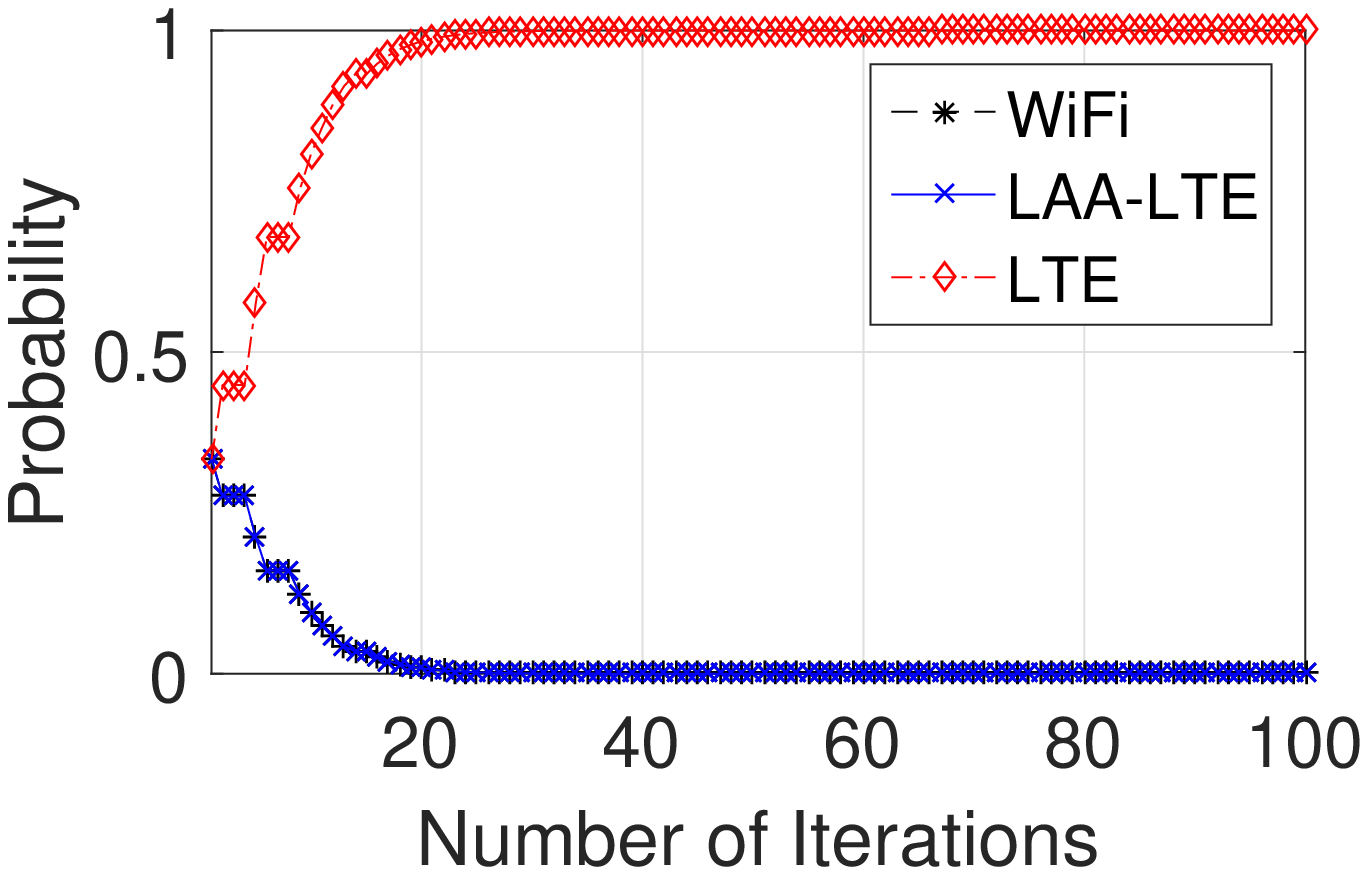}
    \end{minipage} \\[0.9mm]
    \begin{minipage}[b]{0.49\linewidth}
        \centering
        \includegraphics[width=4.0cm]{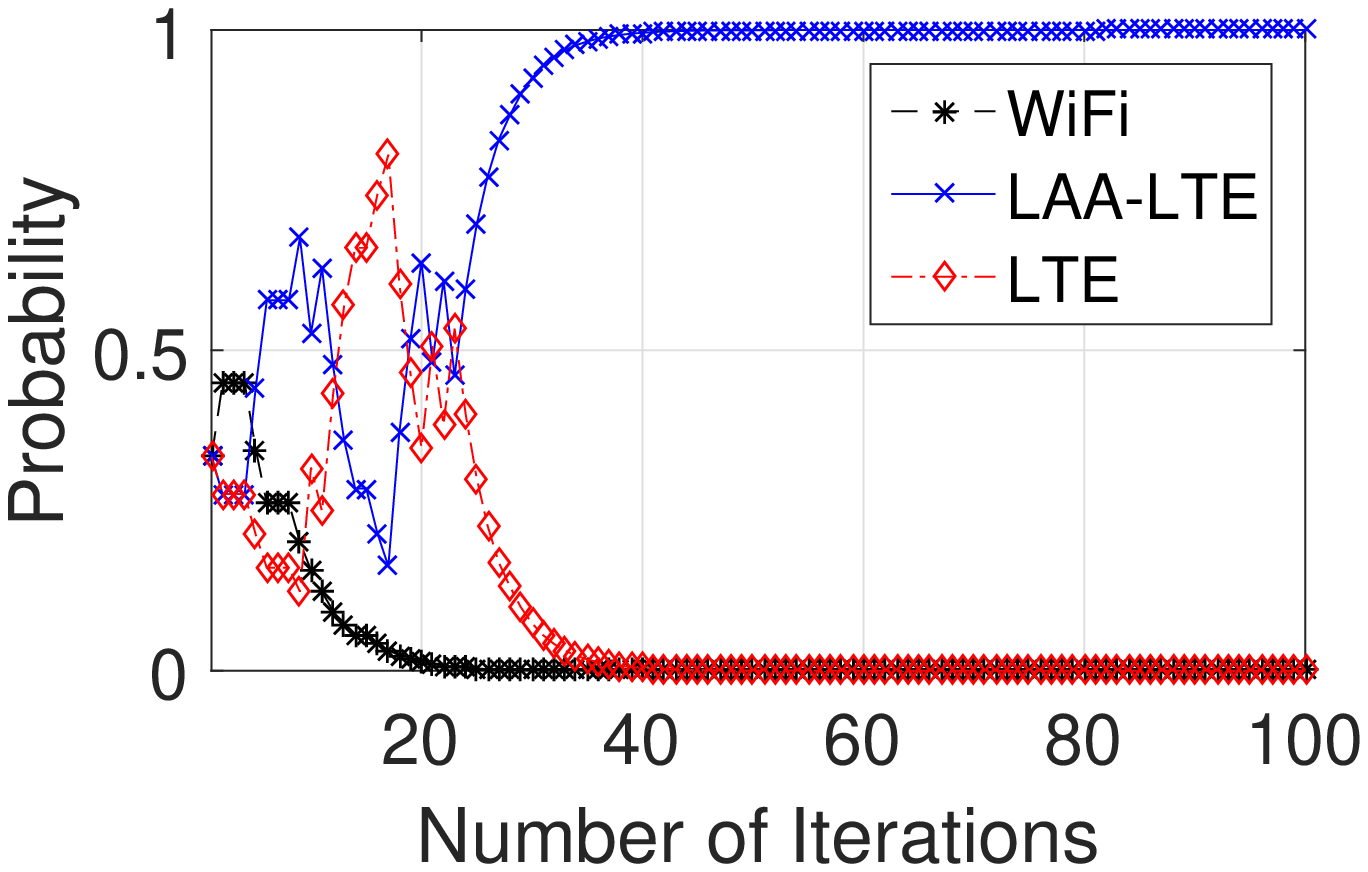}
    \end{minipage}
    \begin{minipage}[b]{0.45\linewidth}
        \centering
        \includegraphics[width=4.0cm]{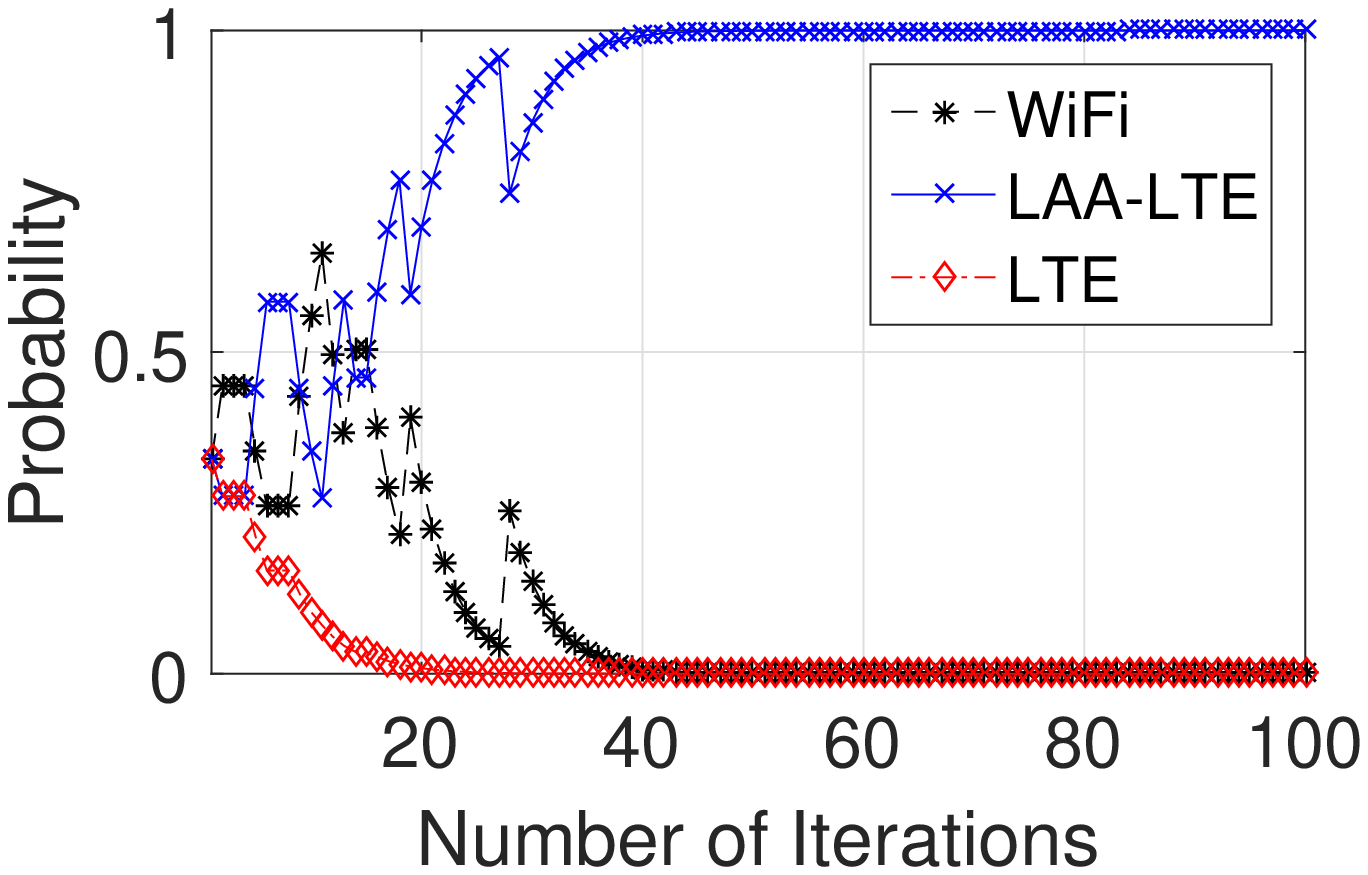}
    \end{minipage}
    \caption{The evolution of mixed strategies of 6 SUs.}
    \vspace{-5pt}
    \label{convergence}
    \vspace{-5pt}
\end{figure}

Fig.\ref{convergence} illustrates the evolution of the mixed strategies of the SUs when NE 3 is finally achieved. It is shown that the SL algorithm converges to a pure NE within tens of iterations, which proves the efficiency of the algorithm.
\begin{table}[t]\label{table1}
\vspace{-10pt}
\centering
\caption{Top 15 Most Frequently Reached NEs and Optimal Points}
\scalebox{0.72}[0.72]{
\begin{tabular}{|c|c|c|c|c|c|c|c|c|}
\hline
 & SU $1$ & SU $2$ & SU $3$ & SU $4$ & SU $5$ & SU $6$ & $R_{total}$ & $\%$\\
\hline
NE1 & WiFi & WiFi & LAA & LTE & LTE & LAA & 0.4636 & $3.656\%$\\
\hline
NE2 & LAA & WiFi & WiFi & LTE & LTE & LAA & 0.4636 & $3.593\%$\\
\hline
NE3$^*$ & WiFi & WiFi & LAA & LTE & LAA & LAA & 0.4754 & $3.357\%$\\
\hline
NE4$^*$ & LAA & WiFi & WiFi & LTE & LAA & LAA & 0.4754 & $3.352\%$\\
\hline
NE5 & WiFi & LTE & LAA & LTE & WiFi & LAA & 0.4593 & $3.151\%$\\
\hline
NE6 & LAA & LTE & WiFi & LTE & WiFi & LAA & 0.4593 & $3.044\%$\\
\hline
NE7 & WiFi & LAA & WiFi & LTE & LAA & LAA & 0.4716 & $2.897\%$\\
\hline
NE8 & WiFi & LAA & LAA & LTE & WiFi & LAA & 0.4711 & $2.871\%$\\
\hline
NE9 & LAA & LAA & LAA & LTE & LAA & WiFi & 0.4716 & $2.863\%$\\
\hline
NE10 & LAA & WiFi & WiFi & LTE & WiFi & LAA & 0.4716 & $2.846\%$\\
\hline
NE11 & LAA & LAA & WiFi & LTE & WiFi & LAA & 0.4711 & $2.802\%$\\
\hline
NE12 & WiFi & WiFi & LAA & LTE & WiFi & LAA & 0.4716 & $2.785\%$\\
\hline
NE13 & WiFi & LAA & WiFi & LTE & LTE & LAA & 0.4516 & $2.315\%$\\
\hline
NE14 & LTE & WiFi & LAA & LTE & WiFi & LAA & 0.4506 & $2.170\%$\\
\hline
NE15 & LAA & WiFi & LTE & LTE & WiFi & LAA & 0.4506 & $2.160\%$\\
\hline
\end{tabular}}
\vspace{-5pt}
\end{table}

\subsection{The Two-Level Learning-Based Mechanism}

In this subsection, we evaluate the performance of the proposed two-level learning-based mechanism in a variational environment. More specifically, there are $5$ IUs and $10$ SUs in the system with $\bm{\lambda}_1=\{$0.03, 0.05, 0.08, 0.09, 0.11$\}$ and $\bm{\lambda}_2=\{$0.01, 0.02, 0.03, 0.03, 0.04, 0.05, 0.08, 0.09, 0.2, 0.3$\}$ at the beginning of the iteration. After a certain time, the numbers of IUs and SUs are changed to 20 and 5, respectively, with the corresponding traffic sets $\bm{\lambda}_1=\{$0.03, 0.04, 0.05, 0.06, 0.06, 0.07, 0.07, 0.03, 0.04, 0.05, 0.06, 0.06, 0.07, 0.07, 0.1, 0.1, 0.2, 0.2, 0.2, 0.2$\}$ and $\bm{\lambda}_2=\{$0.07, 0.08, 0.08, 0.1, 0.2$\}$. The parameters of the proposed eSLQL algorithm are given by $\mathcal{A}_L=\{$0.1, 0.2,$...$, 9.9$\}$, $\alpha=0.1$ and $\delta=5$.

\begin{figure}[t]
\centering
  \includegraphics[width=0.8\columnwidth]{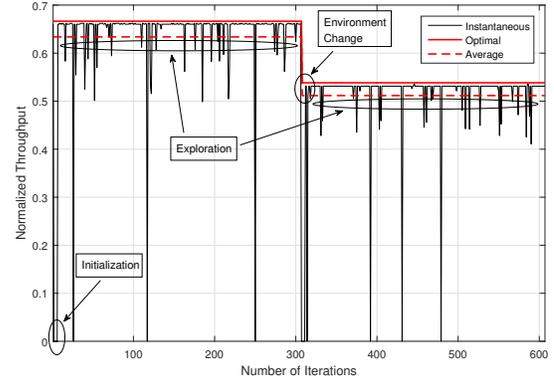}
  \vspace{-5pt}
  \caption{The performance of the proposed two-level learning-based mechanism vs optimal solution.}\label{eslql}
    \vspace{-5pt}
\end{figure}

Fig. \ref{eslql} compares the performance of the proposed learning-based solution with that of the optimal method mentioned in Section III-B. As is depicted in the figure, the proposed algorithm quickly approximates to the optimal performance after the initialization and then takes an immediate action to the variation of the network setting. The huge performance fluctuations in the figure are caused by the exploration mode in the eSLQL algorithm, and the minor ones are induced by the multiple local optimal NEs obtained by the SL algorithm. Though some performance fluctuations exist, the average throughput of the proposed algorithm yields over $95\%$ throughput of the optimal solution in both network settings. The effectiveness and adaptability of the proposed solution are therefore confirmed. In the actual deployment, after the initialization, the iteration can be slowed down to reduce the complexity and the decisions of the previous iteration can be resumed when huge fluctuations are met, to avoid the deep downgrade.

\section{Conclusions}

This article has presented a learning-based coexistence mechanism for LAA-LTE based HetNets. Aiming to maximize the normalized throughput of the unlicensed band while guaranteeing the QoS of users, we have considered the joint resource allocation and network access problem. The two-level framework has been developed to decompose the problem into two subproblems. And then learning-based solutions have been proposed to solve them one by one. The simulation results have shown the proposed solution has achieved near-optimal performance and been more efficient and adaptive due to its distributed and learning-based manner.

\section*{Acknowledgment}
The work is supported by National Natural Science Foundation of China under Grants 61571100, 61631005, 61601247 and 61628103.

\bibliographystyle{IEEEtran}
\bibliography{ref_LAALTE}
\end{document}